\documentclass[prb,final,twocolumn,superscriptaddress,showpacs,showdate]{revtex4}
\usepackage{graphicx}

\begin{document}

\title{Magnetic field dependent specific heat and enhanced Wilson ratio in strongly correlated layered cobalt oxide}

\author{P. Limelette}
\affiliation{Universit\'e Fran\c{c}ois Rabelais, Laboratoire LEMA, UMR 6157 CNRS-CEA, Parc de Grandmont, 37200 Tours, France}
\author{H. Muguerra}
\affiliation{Inorganic Materials Chemistry, Dept. of Chemistry, University of Li\`ege, All\'ee de la Chimie 3 (Bat. B6), 4000 Li\`ege, Belgium}
\author{S. H\'ebert}
\affiliation{Laboratoire CRISMAT, UMR 6508 CNRS-ENSICAEN et Universit\'e de Caen, 6, Boulevard du Mar\'echal Juin, 14050 CAEN Cedex, France}


\begin{abstract}
\vspace{0.3cm}
We have investigated the low temperature specific heat properties as a function of magnetic field in the strongly correlated layered cobalt oxide [BiBa$_{0.66}$K$_{0.36}$O$_2$]CoO$_2$.
These measurements reveal two kinds of magnetic field dependent contributions in qualitative agreement with the presence of a previously inferred  magnetic Quantum Critical Point (QCP).
First, the coefficient of the low temperature T$^3$ behavior of the specific heat turns out to sizeably decrease near a magnetic field consistent with the critical value reported in a recent paper.
In addition, a moderate but significant enhancement of the Sommerfeld coefficient is found in the vicinity of the QCP suggesting a slight  increase of the electronic effective mass.
This result contrasts with the divergent behavior of the previously reported Pauli susceptibility.
Thus, a strongly enhanced Wilson ratio is deduced, suggesting efficient ferromagnetic fluctuations in the Fermi liquid regime which could explain the unusual magnetic field dependent specific heat.
As a strong check, the high magnetic field Wilson ratio asymptotically recovers the universal limit of the local Fermi liquid against ferromagnetism.
\end{abstract}


\maketitle
Transition metal oxides have demonstrated over the last decades how the strong correlations could lead to unanticipated electronic properties. 
Outstanding examples \cite{Imada1998} are superconducting cuprates, manganites with their colossal negative magnetoresistance \cite{Salamon2001}, vanadates displaying the Mott metal-insulator transition \cite{Limelette2003} and the layered cobalt oxides which exhibit an unexpected large thermopower at room temperature \cite{Terasaki1997}.
Most of these oxides share in common that they are doped Mott insulator, \textit{i.e.} their metallicity originates from the introduction of charge carriers by doping, otherwise the strong Coulomb repulsion would localize electrons to form a Mott insulating state \cite{Imada1998,Georges1996}.
Belonging to this class of materials the layered cobalt oxides have revealed, besides their enhanced room temperature thermopower \cite{Wang2003}, a very rich phase diagram as well as striking properties \cite{Bobroff2007,Schulze2008,Nicolaou2010} including large negative magnetoresistance in some compounds \cite{Limelette2008}, or giant electron-electron scattering in Na$_{0.7}$CoO$_2$ \cite{Li2004}. 
Interestingly, the latter observation have already led to conjecture a possible influence of a magnetic QCP in the aforementioned compound. 
Density functional calculations have also predicted at the local spin-density approximation level weak itinerant ferromagnetic state competing with weak itinerant antiferromagnetic state, favoring then quantum critical fluctuations \cite{Singh2007}.

Within this context, susceptibility measurements have recently demonstrated in the strongly correlated layered cobalt oxide [BiBa$_{0.66}$K$_{0.36}$O$_2$]CoO$_2$ the existence of a magnetic quantum critical point governing the electronic properties \cite{Limelette2010}. 
The investigated susceptibility $\chi$ have revealed a scaling behavior with both the temperature T and the magnetic field B ranging from a high-T non-Fermi liquid down to a low-T Fermi liquid.
In the latter Fermi Liquid regime, the Pauli susceptibility has exhibited a divergent behavior with a power law dependence as $\chi \propto$ b$^{-0.6}$ with b=B-B$_C$ which measures the distance from the QCP and the critical magnetic field B$_C \approx$0.176 T.
While several scenarios could explain this result, this behavior may in particular originate from either an enhancement of the electronic effective mass due to the vicinity of the QCP or because of the presence of efficient ferromagnetic fluctuations increasing the Pauli susceptibility by a Stoner factor.
In order to put these scenarios under experimental test, we have investigated the low temperature specific heat properties as a function of magnetic field in the layered cobalt oxide [BiBa$_{0.66}$K$_{0.36}$O$_2$]CoO$_2$ which we report on in this article.

Similarly to Na$_{x}$CoO$_2$, the structure of the layered cobalt oxide [BiBa$_{0.66}$K$_{0.36}$O$_{2}$]CoO$_{2}$ (abbreviated thereafter BBCO) contains single [CoO$_2$] layer of CdI$_2$ type stacked with four rocksalt-type layers, instead of a sodium deficient layer, which act as a charge reservoir \cite{Hervieu2003}.
The reported measurements have been performed on a single-crystal with a mass of 35 mg which was grown using standard flux method \cite{Leligny1999}.
The specific heat has been determined with a calorimeter of a Quantum Design physical properties measurement system using a relaxation method with a temperature rise of the order of 2$\%$ of the sample temperature.
It is worthy to note that the calorimeter, including the thermometers have been calibrated with each magnetic field up to 9 T.
In addition, all the measurements have been duplicated without the sample for each magnetic field in order to compensate both the temperature and magnetic field dependences of the grease used to ensure a good thermal contact between the sample and the calorimeter platform.
We also mention that the calorimeter's parameter which indicates the quality of the thermal contact between the sample and the platform, the so-called sample coupling, has remained between 97$\%$ and 100$\%$ during each measurement below 100 K, ensuring thus the reliability of the results.

\begin{figure}[htbp]
\centerline{\includegraphics[width=0.95\hsize]{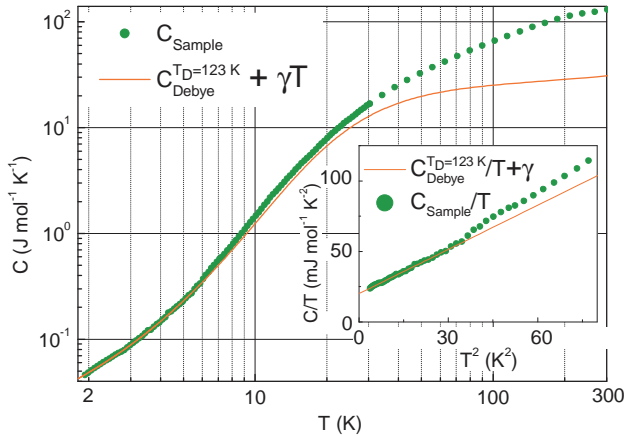}}
\caption{(Color online) Temperature dependence of the single-crystal specific heat compared with a simulation of the specific heat including an electronic part as C$_{el}$=$\gamma$T, and a part due to phonons resulting from the numerical integration of Eq.\ref{eq-debye}.
The former is caracterized by the Sommerfeld coefficient $\gamma \approx$20.1 mJ mol$^{-1}$K$^{-2}$ and the latter by the Debye temperature T$_D$=123 K.
The inset exhibits the low temperature linear regime of C/T as a function of T$^2$. 
}
\label{fig1}
\end{figure}

Figure \ref{fig1} displays the temperature dependence of the specific heat over the range from 300 K down to 1.9 K on a double logarithmic scale in order to show both the low and the high temperature behaviors.
As frequently observed, the main part of the specific heat originates from the acoustic and the optical phonons.
While the latters mainly contribute at the highest temperatures, the formers lead to the usual dependence as C$_{Debye}^{T_D}\approx \beta$ T$^3$ which results from the low temperature expansion of the following equation according to the Debye model.

\begin{equation}
C_{Debye}^{T_D}= 9 N_{Av} k_B \left( \frac{T}{T_D} \right)^3 \int_0^{\frac{T}{T_D}} \frac{x^4 e^x}{(e^x-1)^2} dx
\label{eq-debye}
\end{equation}

Here, N$_{Av}$ is the Avogadro's number, k$_B$ is the Boltzmann constant and T$_D$ the Debye temperature which can be related to the low temperature T$^3$ behavior as $\beta$=12$\pi^4$N$_{Av}$k$_B$/5T$_D^3$.
In addition to the phonons contribution, the electronic part of the specific heat is expected to vary linearly with temperature as C$_{el}$=$\gamma$T, with the Sommerfeld coefficient $\gamma$, as long as T is lower than the Fermi temperature T$_F$.
By integrating numerically Eq. \ref{eq-debye}, the two aforementioned contributions are found to account for the low temperature dependence of the measured specific heat in Fig. \ref{fig1} with $\gamma \approx$20.1 mJ mol$^{-1}$K$^{-2}$ and T$_D$=123 K.
Since the Sommerfeld coefficient is proprotional to the electronic effective mass, the previous rather large value illustrates the presence of sizeable electronic correlations renormalizing the latter mass as observed in several parent compounds.

On the other hand, the inset in Fig. \ref{fig1} reveals the linear regime of C/T as a function of T$^2$ below 6 K with the corresponding coefficient $\beta \approx$1.05 mJ mol$^{-1}$K$^{-4}$ according to T$_D$.
It must here be emphasized that the found parameters are consistent with the previously reported speficic heat data measured in a polycrystalline parent compound \cite{Hervieu2003}.
Interestingly, the inset also displays what could be a sign of an anomaly around 6-7 K in analogy with the one observed quite recently in susceptibility measurements \cite{Limelette2010}.
Nevertheless, its magnitude is not enough pronounced to be firmly ascribed to any electronic instability and thus, this requires more experimental investigations in order to precise its meaning.

\begin{figure}[htbp]
\centerline{\includegraphics[width=0.95\hsize]{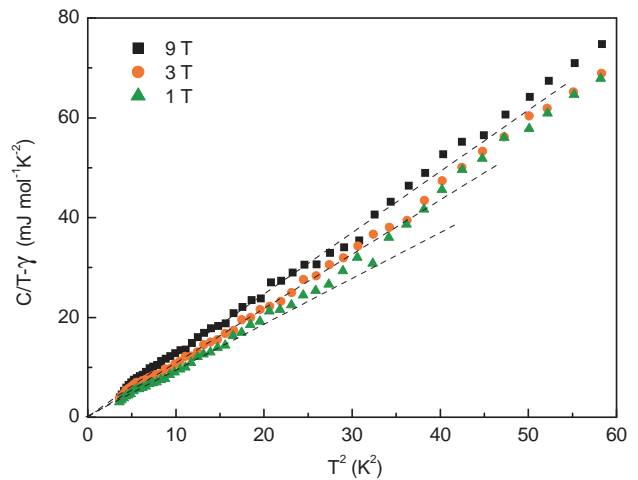}}
\caption{(Color online) Low temperature dependences of the single-crystal specific heat as C/T-$\gamma$ for selected values of the magnetic field spanning the investigated range.
Note that magnetic field dependent Sommerfeld coefficient has been substracted to C/T in order to isolate the variations of the slope.
The dotted lines are guides to the eyes.}
\label{fig2}
\end{figure}

For now, let us turn to the magnetic field dependences of the low temperature specific heat.
The magnetic field has been applied perpendiculary to the CoO$_2$ planes with a magnitude lying within the range up to 9 T.
As displayed in Fig. \ref{fig2} by substracting magnetic field dependent Sommerfeld coefficient to C/T as a function of T$^2$, it turns out that the slope significantly decreases when the field is lowered from 9 T down to 1 T.
Concomitantly, the temperature interval associated with the linear regime follows also this reduction suggesting then a possible magnetic field dependent cross-over.

Since the phonons part of the specific heat could hardly explain such a magnetic field dependence, this result likely suggests another  contribution of electronic type.
Such a behavior could at least be ascribed to two kinds of phenomenon leading to similar temperature dependences than the phonons, namely with a T$^2$ power law for C/T but including nonanalyticy and with a negative sign \cite{Baym1991}.
Therefore, the linear T$^2$ regimes observed in Fig. \ref{fig2} could result from both the standard phonons contribution and another part magnetic field dependent which here reduces the slope as the field is lowered from 9 T down 1 T.

The first scenario involves the finite temperature correction to Fermi liquid theory due to strong interactions between electrons and predicts a temperature dependent Sommerfeld coefficient as $\gamma$(T)$-\gamma \propto $-g$_3$T$^2$ln(T), where g$_3$ is a peculiar coefficient in three dimensions \cite{Chubukov2006,L¨ohneysen2007}, with the zero temperature Sommerfeld coefficient $\gamma$.
Since the magnitude of this correction is of the order of (T/T$_F$)$^2$, it could increase if the Fermi temperature decreases with the magnetic field.
Within the frame of this interpretation, a decrease of the Fermi temperature could result from an enhancement of the electronic correlations.
Therefore, the latter should implies an increase of the effective mass and the Sommerfeld coefficient should follow this enhancement too.

The other scenario is concerned with the paramagnon model used to describe itinerant electrons near ferromagnetic instability \cite{Doniach1966,Chubukov2006}.
Due to some formal analogies between the analysis realised within the finite temperature corrected Fermi liquid theory and the paramagnon model, a similar temperature dependence as $\gamma$(T)$-\gamma \propto $-g$_{spin}$T$^2$ln(T) is found, with the corresponding coefficient g$_{spin}$.
In contrast to the first scenario, the latter model involves electronic spin-spin interactions and leads to a coefficient g$_{spin}$  proportional to a Stoner factor potentially enhanced in the vicinity of a ferromagnetic instability.
It is worthmentioning that these kinds of correction to the specific heat have been widely discussed in the contexts of liquid $^3$He \cite{Levin1983,Vollhardt1984} and also heavy fermions compounds \cite{Visser1987}.

\begin{figure}[htbp]
\centerline{\includegraphics[width=0.95\hsize]{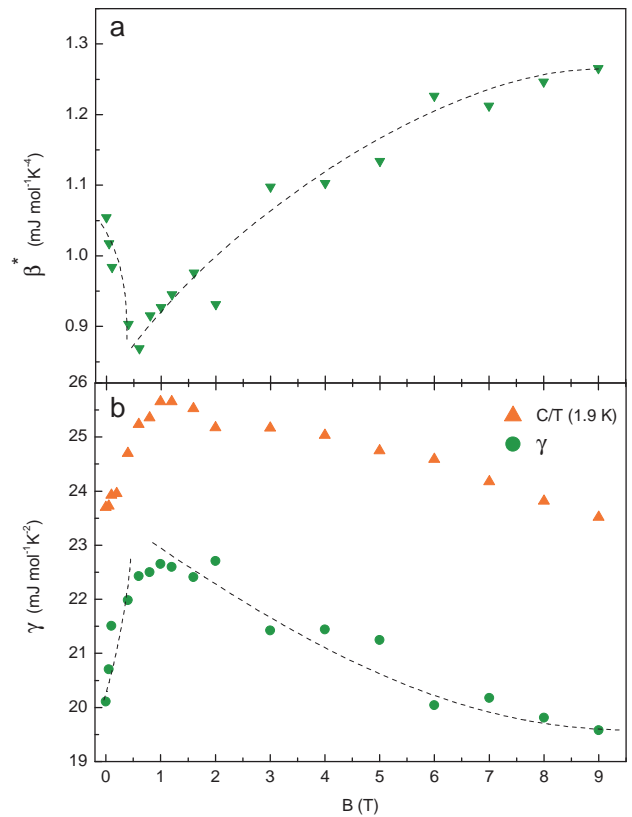}}
\caption{(Color online) Magnetic field dependences of (a) the effective slope $\beta^*$ and (b) the Sommerfeld coefficient $\gamma$ which is compared with the specific heat as C/T measured at the lowest temperature (1.9 K).
The dotted lines are guides to the eyes.}
\label{fig3}
\end{figure}

To further proceed the analysis, the effective slope of the T$^2$ regime, namely $\beta^*$, and the Sommerfeld coefficients have been systematically determined within the investigated magnetic field range and are reported in Fig. \ref{fig3}.
As a result, Fig. \ref{fig3}a reveals a non monotonic magnetic field dependence of $\beta^*$ with first a decrease down to nearly 0.5 T followed by an increase with B as already exemplified in Fig. \ref{fig2}.
The magnetic field dependence of $\gamma$ exactly displays in Fig. \ref{fig3}b the complementary behavior with a broad maximum around 0.5 T, instead of the minimum of $\beta^*$.
Since these two parameters are not determined independently, the latter figure also shows the experimental values of C/T measured at the lowest temperature.
As a strong check, the magnetic field dependence of these experimental values is qualitatively the same than for $\gamma$ ensuring thus that these behaviors originate from a physical effect and are not an artefact of the procedure.
One must here emphasize that the observed variations are rather moderate, especially those of $\gamma$ which could be a sign of a logarithmic magnetic field dependence as expected in the case of a nearly ferromagnetic Fermi liquid \cite{Doniach1966}.
Note that such a weak renormalization is also theoretically predicted in the vicinity of a magnetic quantum critical point within some scenarios of the Hertz-Millis theory \cite{Hertz1976,Millis1993}.
Anyway, Fig. \ref{fig3}b suggests a slight enhancement of the electronic effective mass near a magnetic field which is roughly consistent with the critical value determined from susceptibility measurements \cite{Limelette2010}.
Therefore, the behaviors of $\gamma$ and $\beta^*$ seem to be related to the magnetic quantum critical point inferred from susceptibility measurements.

\begin{figure}[htbp]
\centerline{\includegraphics[width=0.95\hsize]{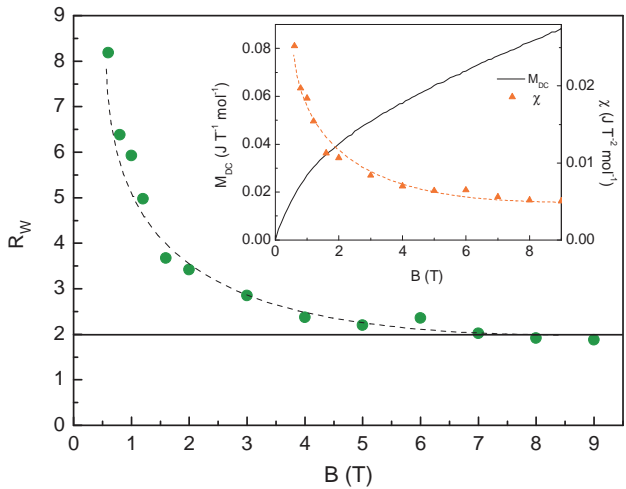}}
\caption{(Color online) Magnetic field dependence of the dimensionless Wilson Ratio R$_W$=$ \pi^2/3 \left( k_B/\mu_B \right)^2 \chi/\gamma$.
The inset displays the magnetic field dependences of both the magnetization M$_{DC}$ measured at 1.9 K (left axis) and the susceptibility $\chi$ (right axis) inferred from a numerical differentiation of M$_{DC}$ and used in the Wilson Ratio.
Note that the magnetic field has been applied perpendiculary to the CoO$_2$ planes.
The dotted lines are guides to the eyes.
}
\label{fig4}
\end{figure}

In order to discriminate these scenarios, let's combine the specific heat results with the previously reported Pauli susceptibility.
Indeed, within the Landau's Fermi liquid theory the electronic correlations renormalize the quasiparticles effective mass, and then enhance both the Pauli susceptibility and the Sommerfeld coefficient.
For a non-interacting Fermi gas, the former quantity writes $\chi_0$=$\mu_B^2 g_F$ and the latter $\gamma_0$=$(\pi^2/3) k_B^2 g_F$, with the Bohr magneton $\mu_B$ and the density of states at the Fermi level $g_F$.
One can therefore define a dimensionless quantity known as the Wilson ratio R$_W$ \cite{Wilson1975}:
$$
R_W= \frac{\pi^2 k_B^2}{3 \mu_B^2} \frac{\chi}{\gamma}
$$
Thus, it immediately follows that R$_W$= 1 for the free electrons.
Here, by using the determined values of both the Pauli susceptibility and the Sommerfeld coefficient as respectively displayed in the inset of Fig. $\ref{fig4}$ and Fig. $\ref{fig3}$b, the Wilson ratio is found to reveal in Fig. $\ref{fig4}$ a strong enhancement up to R$_W \approx$ 8 at B = 0.5 T, namely in the vicinity of the QCP.

In addition, the values of this ratio seem to saturate above B$\approx$7 T down to R$_W$=2 and appear thus, always higher than the free electrons one.
To interpret these features, it is now instructive to give a more precise insight into the Landau Fermi liquid theory.
Besides the dimensionless Fermi liquid mass enhancement $m^*$ concerning both $\chi \propto \chi_0 \: m^*$ and $\gamma = \gamma_0 \: m^*$, the Pauli susceptibility in equation \ref{eq7} is additionally renormalized by one of the first Landau parameters $F_0^a$ as the Wilson ratio accordingly \cite{Pines1966,Vollhardt1984}.
\begin{equation}
\chi= \chi_0 \frac{m^*}{1+F_0^a}  \mbox{\ \ \ $\Longrightarrow$  \ \ \ }  R_W= \frac{1}{1+F_0^a} 
\label{eq7}
\end{equation}
Whereas the Landau parameters can hardly be theoretically predicted, some of their asymptotic behaviors can be inferred in the case of the Local Fermi Liquid (LFL) \cite{Engelbrecht1995}.
Characterized by a momentum independent self energy, the LFL allows to recover some of the essential features of the Kondo model described by Wilson \cite{Wilson1975} as well as some of the Fermi liquid properties obtained within the framework of the dynamical mean field theory \cite{Georges1996}.

In the Landau theory, one can relate the Landau parameters $F^{s,a}_l$ to the scattering amplitudes $A^{s,a}_l=F^{s,a}_l/(1+F^{s,a}_l/(2l+1))$ by performing a standard spherical harmonic analysis, where the upperscripts s and a respectively refer to the spin-symmetric and the spin-antisymmetric parts.
Also, because of the momentum independence of the LFL self energy, the only non vanishing aforementioned parameters are of s-wave type, namely with l=0.
It results that the forward scattering sum rule $\Sigma_l \left( A^s_l+A^a_l \right)= \: 0$ simplifies to $A^s_0=-A^a_0$.
Thus, by considering the unitarity limit with the scattering amplitude $A^s_0 \rightarrow 1$, one deduces the asymptotic behaviors for the remaining Landau parameters $F^s_0 \rightarrow \infty$ and $F^a_0 \rightarrow -1/2$.
Following equation \ref{eq7}, the Wilson ratio lies between 1, for the free electrons limit, and 2 as long as the LFL is robust against ferromagnetic or metal insulator instabilities \cite{Engelbrecht1995}.
Interestingly, the latter universal limit R$_W$=2 appears to be achieved in Fig. $\ref{fig4}$, suggesting then that a Local Fermi Liquid regime is reached in this system above B$\approx$7 T.
Since the susceptibility remains Pauli-like within the Fermi Liquid regime, one infers efficient ferromagnetic fluctuations which destabilize the LFL regime below B$\approx$7 T and strongly increase the susceptibility values due to a Stoner-like enhancement \cite{Vollhardt1984} in the vicinity of the QCP.
Let's finally emphasize that besides the close analogy between the scaling properties of the susceptibility measured in this compound \cite{Limelette2010} and in the heavy fermion system YbRh$_2$(Si$_{0.95}$Ge$_{0.05}$)$_2$ \cite{Gegenwart2005}, the found Wilson ratios  are also qualitatively identical.
In both system, R$_W$ displays a strong enhancement in the vicinity of the QCP and recovers a constant value at high magnetic field which however numerically differs according to the system.
In particular, the Wilson ratio can reach values as high as 17.5 in the aforementioned heavy fermion compound \cite{Gegenwart2005}, 3-4 in CeCu$_6$ and CeRu$_2$Si$_2$ \cite{krug2003}, 10 in Sr$_3$Ru$_3$O$_7$ \cite{ikeda2000}, 6-8 (Pd), 12 (TiBe$_2$) or even 40 (Ni$_3$Ga) in nearly ferromagnetic metals \cite{julian1999}.

As a result, the values of the Wilson ratio are clearly in favor of a magnetic field induced nearly ferromagnetic Fermi liquid.
They also indicate that the high magnetic field dependence of the specific heat coefficient $\beta^*$ in Fig. \ref{fig3}a can be mainly ascribed to ferromagnetic spin fluctuations.
Moreover, both the magnetic field dependences of $\beta^*$ and $\gamma$ are qualitatively consistent with the presence of a QCP.
The latter could therefore originate from a competition between ferromagnetic fluctuations and antiferromagnetic interactions at low magnetic field as assumed from susceptibility measurements \cite{Limelette2010}.
Further experimental investigations are now required in order to precisely characterized the complex low magnetic field regime.

\begin{acknowledgments}
We are grateful to W. Saulquin for usefull discussions and we would like to acknowledge support from La R\'egion Centre.
\end{acknowledgments}


\begin{thebibliography}{20}
\expandafter\ifx\csname natexlab\endcsname\relax\def\natexlab#1{#1}\fi
\expandafter\ifx\csname bibnamefont\endcsname\relax
  \def\bibnamefont#1{#1}\fi
\expandafter\ifx\csname bibfnamefont\endcsname\relax
  \def\bibfnamefont#1{#1}\fi
\expandafter\ifx\csname citenamefont\endcsname\relax
  \def\citenamefont#1{#1}\fi
\expandafter\ifx\csname url\endcsname\relax
  \def\url#1{\texttt{#1}}\fi
\expandafter\ifx\csname urlprefix\endcsname\relax\def\urlprefix{URL }\fi
\providecommand{\bibinfo}[2]{#2}
\providecommand{\eprint}[2][]{\url{#2}}


\bibitem[{\citenamefont{Imada et~al.}(1998)\citenamefont{Imada}}]{Imada1998}
\bibinfo{author}{\bibfnamefont{M.}~\bibnamefont{Imada}}, 
\bibinfo{author}{\bibfnamefont{A.}~\bibnamefont{Fujimori}} \bibnamefont{and}
\bibinfo{author}{\bibfnamefont{Y.}~\bibnamefont{Tokura}}, 
\bibinfo{journal}{Rev. Mod. Phys.} \textbf{\bibinfo{volume}{70}},
\bibinfo{pages}{1039} (\bibinfo{year}{1998}).
  
  
  \bibitem[{\citenamefont{Salamon et~al.}(2001)\citenamefont{Salamon}}]{Salamon2001}
\bibinfo{author}{\bibfnamefont{M.B.}~\bibnamefont{Salamon}} \bibnamefont{and}
  \bibinfo{author}{\bibfnamefont{M.}~\bibnamefont{Jaime}},
  \bibinfo{journal}{Rev. Mod. Phys.} \textbf{\bibinfo{volume}{73}},
  \bibinfo{pages}{583} (\bibinfo{year}{2001}).
  
\bibitem[{\citenamefont{Limelette et~al.}(2003)\citenamefont{Limelette}}]{Limelette2003}
\bibinfo{author}{\bibfnamefont{P.}~\bibnamefont{Limelette}}, 
\bibinfo{author}{\bibfnamefont{A.}~\bibnamefont{Georges}}, 
\bibinfo{author}{\bibfnamefont{P.}~\bibnamefont{Wzietek}}, 
\bibinfo{author}{\bibfnamefont{D.}~\bibnamefont{Jerome}}, 
\bibinfo{author}{\bibfnamefont{P.}~\bibnamefont{Metcalf}} \bibnamefont{and}
\bibinfo{author}{\bibfnamefont{J.}~\bibnamefont{Honig}}, 
\bibinfo{journal}{Science} \textbf{\bibinfo{volume}{302}},
\bibinfo{pages}{89} (\bibinfo{year}{2003});
\bibinfo{author}{\bibfnamefont{S.}~\bibnamefont{Autier-Laurent}}, 
\bibinfo{author}{\bibfnamefont{B.}~\bibnamefont{Mercey}}, 
\bibinfo{author}{\bibfnamefont{D.}~\bibnamefont{Chippaux}}, 
\bibinfo{author}{\bibfnamefont{P.}~\bibnamefont{Limelette}} \bibnamefont{and}
\bibinfo{author}{\bibfnamefont{Ch.}~\bibnamefont{Simon}},
  \bibinfo{journal}{Phys. Rev. B} \textbf{\bibinfo{volume}{74}},
  \bibinfo{pages}{195109} (\bibinfo{year}{2006}).
  
  \bibitem[{\citenamefont{Terasaki et~al.}(1997)\citenamefont{Terasaki, Sasago and Uchinokura}}]{Terasaki1997}
\bibinfo{author}{\bibfnamefont{I.}~\bibnamefont{Terasaki}},
  \bibinfo{author}{\bibfnamefont{Y.}~\bibnamefont{Sasago}} \bibnamefont{and}
  \bibinfo{author}{\bibfnamefont{K.}~\bibnamefont{Uchinokura}}, 
  \bibinfo{journal}{Phys. Rev. B} \textbf{\bibinfo{volume}{56}},
  \bibinfo{pages}{R12685} (\bibinfo{year}{1997}).


\bibitem[{\citenamefont{Georges et~al.}(1996)\citenamefont{Georges, Kotliar,
  Krauth, and Rozenberg}}]{Georges1996}
\bibinfo{author}{\bibfnamefont{A.}~\bibnamefont{Georges}},
  \bibinfo{author}{\bibfnamefont{G.}~\bibnamefont{Kotliar}},
  \bibinfo{author}{\bibfnamefont{W.}~\bibnamefont{Krauth}} \bibnamefont{and}
  \bibinfo{author}{\bibfnamefont{M.J.} \bibnamefont{Rozenberg}}, 
  \bibinfo{journal}{Rev. Mod. Phys.} \textbf{\bibinfo{volume}{68}},
  \bibinfo{pages}{13} (\bibinfo{year}{1996}).


\bibitem[{\citenamefont{Wang}(2003)\citenamefont{Wang}}]{Wang2003}
\bibinfo{author}{\bibfnamefont{Y.}~\bibnamefont{Wang}},
\bibinfo{author}{\bibfnamefont{N.S.}~\bibnamefont{Rogado}},
\bibinfo{author}{\bibfnamefont{R.J.}~\bibnamefont{Cava}} \bibnamefont{and}
\bibinfo{author}{\bibfnamefont{N.P.}~\bibnamefont{Ong}}, 
\bibinfo{journal}{Nature} \textbf{\bibinfo{volume}{423}},
\bibinfo{pages}{425} (\bibinfo{year}{2003});
\bibinfo{author}{\bibfnamefont{P.}~\bibnamefont{Limelette}},
\bibinfo{author}{\bibfnamefont{S.}~\bibnamefont{H\'ebert}}, 
\bibinfo{author}{\bibfnamefont{V.}~\bibnamefont{Hardy}}, 
\bibinfo{author}{\bibfnamefont{R.}~\bibnamefont{Fr\'esard}}, 
\bibinfo{author}{\bibfnamefont{Ch.}~\bibnamefont{Simon}} \bibnamefont{and}
\bibinfo{author}{\bibfnamefont{A.}~\bibnamefont{Maignan}}, 
\bibinfo{journal}{Phys. Rev. Lett.} \textbf{\bibinfo{volume}{97}},
\bibinfo{pages}{046601} (\bibinfo{year}{2006}).
 
\bibitem[{\citenamefont{Bobroff et~al.}(2007)\citenamefont{Bobroff}}]{Bobroff2007}
\bibinfo{author}{\bibfnamefont{J.}~\bibnamefont{Bobroff}},
\bibinfo{author}{\bibfnamefont{S.}~\bibnamefont{H\'ebert}}, 
\bibinfo{author}{\bibfnamefont{G.}~\bibnamefont{Lang}}, 
\bibinfo{author}{\bibfnamefont{P.}~\bibnamefont{Mendels}}, 
\bibinfo{author}{\bibfnamefont{D.}~\bibnamefont{Pelloquin}} \bibnamefont{and}
\bibinfo{author}{\bibfnamefont{A.}~\bibnamefont{Maignan}}, 
\bibinfo{journal}{Phys. Rev. B} \textbf{\bibinfo{volume}{76}},
\bibinfo{pages}{R100407} (\bibinfo{year}{2007}). 


\bibitem[{\citenamefont{Schulze et~al.}(2008)\citenamefont{Schulze}}]{Schulze2008}
\bibinfo{author}{\bibfnamefont{T.F.}~\bibnamefont{Schulze}},
\bibinfo{author}{\bibfnamefont{M.}~\bibnamefont{Bruhwiler}}, 
\bibinfo{author}{\bibfnamefont{P.S.}~\bibnamefont{Hafliger}}, 
\bibinfo{author}{\bibfnamefont{S.M.}~\bibnamefont{Kazakov}}, 
\bibinfo{author}{\bibfnamefont{Ch.}~\bibnamefont{Niedermayer}},
\bibinfo{author}{\bibfnamefont{K.}~\bibnamefont{Mattenberger}},
\bibinfo{author}{\bibfnamefont{J.}~\bibnamefont{Karpinski}} \bibnamefont{and}
\bibinfo{author}{\bibfnamefont{B.}~\bibnamefont{Batlogg}}, 
\bibinfo{journal}{Phys. Rev. B} \textbf{\bibinfo{volume}{78}},
\bibinfo{pages}{205101} (\bibinfo{year}{2008}). 


\bibitem[{\citenamefont{Nicolaou et~al.}(2010)\citenamefont{Nicolaou}}]{Nicolaou2010}
\bibinfo{author}{\bibfnamefont{A.}~\bibnamefont{Nicolaou}},
\bibinfo{author}{\bibfnamefont{V.}~\bibnamefont{Brouet}}, 
\bibinfo{author}{\bibfnamefont{M.}~\bibnamefont{Zacchigna}}, 
\bibinfo{author}{\bibfnamefont{I.}~\bibnamefont{Vobornik}}, 
\bibinfo{author}{\bibfnamefont{A.}~\bibnamefont{Tejeda}},
\bibinfo{author}{\bibfnamefont{A.}~\bibnamefont{Taleb-Ibrahimi}},
\bibinfo{author}{\bibfnamefont{P.}~\bibnamefont{Le F\`evre}}, 
\bibinfo{author}{\bibfnamefont{F.}~\bibnamefont{Bertran}}, 
\bibinfo{author}{\bibfnamefont{S.}~\bibnamefont{H\'ebert}}, 
\bibinfo{author}{\bibfnamefont{H.}~\bibnamefont{Muguerra}} \bibnamefont{and}
\bibinfo{author}{\bibfnamefont{D.}~\bibnamefont{Grebille}}, 
\bibinfo{journal}{Phys. Rev. Lett.} \textbf{\bibinfo{volume}{104}},
\bibinfo{pages}{056403} (\bibinfo{year}{2010}). 


  
\bibitem[{\citenamefont{Limelette et~al.}(2008)\citenamefont{Limelette}}]{Limelette2008}
\bibinfo{author}{\bibfnamefont{P.}~\bibnamefont{Limelette}},
\bibinfo{author}{\bibfnamefont{J.C.}~\bibnamefont{Soret}}, 
\bibinfo{author}{\bibfnamefont{H.}~\bibnamefont{Muguerra}} \bibnamefont{and}
\bibinfo{author}{\bibfnamefont{D.}~\bibnamefont{Grebille}}, 
\bibinfo{journal}{Phys. Rev. B.} \textbf{\bibinfo{volume}{77}},
\bibinfo{pages}{245123} (\bibinfo{year}{2008}); 
\bibinfo{author}{\bibfnamefont{P.}~\bibnamefont{Limelette}},
\bibinfo{author}{\bibfnamefont{S.}~\bibnamefont{H\'ebert}}, 
\bibinfo{author}{\bibfnamefont{H.}~\bibnamefont{Muguerra}}, 
\bibinfo{author}{\bibfnamefont{R.}~\bibnamefont{Fr\'esard}}, \bibnamefont{and}
\bibinfo{author}{\bibfnamefont{Ch.}~\bibnamefont{Simon}},
\bibinfo{journal}{Phys. Rev. B.} \textbf{\bibinfo{volume}{77}},
\bibinfo{pages}{235118} (\bibinfo{year}{2008}).


\bibitem[{\citenamefont{Li}(2004)\citenamefont{Li}}]{Li2004}
\bibinfo{author}{\bibfnamefont{S.Y.}~\bibnamefont{Li}}, 
\bibinfo{author}{\bibfnamefont{L.}~\bibnamefont{Taillefer}}, 
\bibinfo{author}{\bibfnamefont{D.G.}~\bibnamefont{Hawthorn}}, 
\bibinfo{author}{\bibfnamefont{M.A.}~\bibnamefont{Tanatar}}, 
\bibinfo{author}{\bibfnamefont{J.}~\bibnamefont{Paglione}}, 
\bibinfo{author}{\bibfnamefont{M.}~\bibnamefont{Sutherland}}, 
\bibinfo{author}{\bibfnamefont{R.W.}~\bibnamefont{Hill}}, 
\bibinfo{author}{\bibfnamefont{C.H.}~\bibnamefont{Wang}} and 
\bibinfo{author}{\bibfnamefont{X.H.}~\bibnamefont{Chen}}, 
\bibinfo{journal}{Phys. Rev. Lett.} \textbf{\bibinfo{volume}{93}},
\bibinfo{pages}{056401} (\bibinfo{year}{2004}).
  
 
  
\bibitem[{\citenamefont{Singh}(2003)\citenamefont{Singh}}]{Singh2007}
\bibinfo{author}{\bibfnamefont{D.J.}~\bibnamefont{Singh}}, 
\bibinfo{journal}{Phys. Rev. B} \textbf{\bibinfo{volume}{68}},
\bibinfo{pages}{R020503} (\bibinfo{year}{2003}).

  
\bibitem[{\citenamefont{Limelette et~al.}(2010)\citenamefont{Limelette}}]{Limelette2010}
\bibinfo{author}{\bibfnamefont{P.}~\bibnamefont{Limelette}},
\bibinfo{author}{\bibfnamefont{W.}~\bibnamefont{Saulquin}}, 
\bibinfo{author}{\bibfnamefont{H.}~\bibnamefont{Muguerra}} \bibnamefont{and}
\bibinfo{author}{\bibfnamefont{D.}~\bibnamefont{Grebille}}, 
\bibinfo{journal}{Phys. Rev. B.} \textbf{\bibinfo{volume}{81}},
\bibinfo{pages}{115113} (\bibinfo{year}{2010}).

 
\bibitem[{\citenamefont{Hervieu et~al.}(2003)\citenamefont{Hervieu}}]{Hervieu2003}
\bibinfo{author}{\bibfnamefont{M.}~\bibnamefont{Hervieu}},
\bibinfo{author}{\bibfnamefont{A.}~\bibnamefont{Maignan}}, 
\bibinfo{author}{\bibfnamefont{C.}~\bibnamefont{Michel}}, 
\bibinfo{author}{\bibfnamefont{V.}~\bibnamefont{Hardy}}, 
\bibinfo{author}{\bibfnamefont{N.}~\bibnamefont{Cr\'eon}} \bibnamefont{and}
\bibinfo{author}{\bibfnamefont{B.}~\bibnamefont{Raveau}}, 
\bibinfo{journal}{Phys. Rev. B} \textbf{\bibinfo{volume}{67}},
\bibinfo{pages}{045112} (\bibinfo{year}{2003}).


\bibitem[{\citenamefont{Leligny et~al.}(1999)\citenamefont{Leligny}}]{Leligny1999}
\bibinfo{author}{\bibfnamefont{H.}~\bibnamefont{Leligny}},
\bibinfo{author}{\bibfnamefont{D.}~\bibnamefont{Grebille}}, 
\bibinfo{author}{\bibfnamefont{A.C.}~\bibnamefont{Masset}}, 
\bibinfo{author}{\bibfnamefont{M.}~\bibnamefont{Hervieu}}, 
\bibinfo{author}{\bibfnamefont{C.}~\bibnamefont{Michel}} \bibnamefont{and}
\bibinfo{author}{\bibfnamefont{B.}~\bibnamefont{Raveau}}, 
\bibinfo{journal}{C.R. Acad. Sci., Ser. IIc: Chim} \textbf{\bibinfo{volume}{2}},
\bibinfo{pages}{409} (\bibinfo{year}{1999}).


\bibitem[{\citenamefont{Baym}(1991)\citenamefont{Baym}}]{Baym1991}
\bibinfo{author}{\bibfnamefont{G.}~\bibnamefont{Baym}} \bibnamefont{and}
\bibinfo{author}{\bibfnamefont{C.}~\bibnamefont{Pethick}},
\bibinfo{author}{\bibfnamefont{Landau Fermi-Liquid Theory (Wiley, New York, 1991)}}.



\bibitem[{\citenamefont{Chubukov et~al.}(2006)\citenamefont{Chubukov}}]{Chubukov2006}
\bibinfo{author}{\bibfnamefont{A.V.}~\bibnamefont{Chubukov}}, 
\bibinfo{author}{\bibfnamefont{D.L.}~\bibnamefont{Maslov}} \bibnamefont{and}
\bibinfo{author}{\bibfnamefont{A.J.}~\bibnamefont{Millis}}, 
  \bibinfo{journal}{Phys. Rev. B} \textbf{\bibinfo{volume}{73}},
  \bibinfo{pages}{045128} (\bibinfo{year}{2006}).
  

\bibitem[{\citenamefont{L¨ohneysen et~al.}(2007)\citenamefont{L¨ohneysen}}]{L¨ohneysen2007}
\bibinfo{author}{\bibfnamefont{H.v.}~\bibnamefont{Lohneysen}}, 
\bibinfo{author}{\bibfnamefont{A.}~\bibnamefont{Rosch}}, 
\bibinfo{author}{\bibfnamefont{M.}~\bibnamefont{Vojta}}  \bibnamefont{and}
\bibinfo{author}{\bibfnamefont{P.}~\bibnamefont{W¨olfle}}, 
  \bibinfo{journal}{Rev. Mod. Phys.} \textbf{\bibinfo{volume}{79}},
  \bibinfo{pages}{1015} (\bibinfo{year}{2007}).


\bibitem[{\citenamefont{Doniach et~al.}(2006)\citenamefont{Doniach}}]{Doniach1966}
\bibinfo{author}{\bibfnamefont{S.}~\bibnamefont{Doniach}} \bibnamefont{and}
\bibinfo{author}{\bibfnamefont{S.}~\bibnamefont{Engelsberg}}, 
  \bibinfo{journal}{Phys. Rev. Lett.} \textbf{\bibinfo{volume}{17}},
  \bibinfo{pages}{750} (\bibinfo{year}{1966}).


\bibitem[{\citenamefont{Levin}(1983)\citenamefont{Levin}}]{Levin1983}
\bibinfo{author}{\bibfnamefont{K.}~\bibnamefont{Levin}} \bibnamefont{and}
\bibinfo{author}{\bibfnamefont{O.T.}~\bibnamefont{Valls}},
\bibinfo{journal}{Phys. Rep.} \textbf{\bibinfo{volume}{98}},
\bibinfo{pages}{1} (\bibinfo{year}{1981}).

  
\bibitem[{\citenamefont{Vollhardt}(1984)\citenamefont{Vollhardt}}]{Vollhardt1984}
\bibinfo{author}{\bibfnamefont{D.}~\bibnamefont{Vollhardt}}, 
\bibinfo{journal}{Rev. Mod. Phys.} \textbf{\bibinfo{volume}{56}},
\bibinfo{pages}{99} (\bibinfo{year}{1984}).


  
\bibitem[{\citenamefont{Visser}(1987)\citenamefont{Visser}}]{Visser1987}
\bibinfo{author}{\bibfnamefont{A. de}~\bibnamefont{Visser}}, 
 \bibinfo{author}{\bibfnamefont{A.}~\bibnamefont{Menovsky}} \bibnamefont{and}
\bibinfo{author}{\bibfnamefont{J.J.M.}~\bibnamefont{Franses}},
\bibinfo{journal}{Physica B} \textbf{\bibinfo{volume}{147}},
\bibinfo{pages}{81} (\bibinfo{year}{1987}).


\bibitem[{\citenamefont{Hertz}(1976)\citenamefont{Hertz}}]{Hertz1976}
  \bibinfo{author}{\bibfnamefont{J.A.}~\bibnamefont{Hertz}}, 
  \bibinfo{journal}{Phys. Rev. B} \textbf{\bibinfo{volume}{14}},
  \bibinfo{pages}{1165} (\bibinfo{year}{1976}).
  
  
\bibitem[{\citenamefont{Millis}(1993)\citenamefont{Millis}}]{Millis1993}
 \bibinfo{author}{\bibfnamefont{A.J.}~\bibnamefont{Millis}}, 
  \bibinfo{journal}{Phys. Rev. B} \textbf{\bibinfo{volume}{48}},
  \bibinfo{pages}{7183} (\bibinfo{year}{1993}).



\bibitem[{\citenamefont{Wilson}(1975)\citenamefont{Wilson}}]{Wilson1975}
  \bibinfo{author}{\bibfnamefont{K.G.}~\bibnamefont{Wilson}}, 
  \bibinfo{journal}{Rev. Mod. Phys.} \textbf{\bibinfo{volume}{47}},
  \bibinfo{pages}{773} (\bibinfo{year}{1975}).

\bibitem[{\citenamefont{Pines}(1966)\citenamefont{Pines}}]{Pines1966}
\bibinfo{author}{\bibfnamefont{D.}~\bibnamefont{Pines}} \bibnamefont{and}
\bibinfo{author}{\bibfnamefont{P.}~\bibnamefont{Nozi\`eres}},
\bibinfo{author}{\bibfnamefont{The Theory of Quantum Liquids (Benjamin, New York, 1966)}}.


\bibitem[{\citenamefont{Engelbrecht et~al.}(1995)\citenamefont{Engelbrecht}}]{Engelbrecht1995}
\bibinfo{author}{\bibfnamefont{J.R.}~\bibnamefont{Engelbrecht}} \bibnamefont{and}
\bibinfo{author}{\bibfnamefont{K.S.}~\bibnamefont{Bedell}}, 
  \bibinfo{journal}{Phys. Rev. Lett.} \textbf{\bibinfo{volume}{74}},
  \bibinfo{pages}{4265} (\bibinfo{year}{1995}).



  
\bibitem[{\citenamefont{Gegenwart et~al.}(2005)\citenamefont{Gegenwart}}]{Gegenwart2005}
\bibinfo{author}{\bibfnamefont{P.}~\bibnamefont{Gegenwart}},
\bibinfo{author}{\bibfnamefont{J.}~\bibnamefont{Custers}},
\bibinfo{author}{\bibfnamefont{Y.}~\bibnamefont{Tokiwa}},
\bibinfo{author}{\bibfnamefont{C.}~\bibnamefont{Geibel}} \bibnamefont{and}
\bibinfo{author}{\bibfnamefont{F.}~\bibnamefont{Steglich}}, 
\bibinfo{journal}{Phys. Rev. Lett.} \textbf{\bibinfo{volume}{94}},
\bibinfo{pages}{076402} (\bibinfo{year}{2005}).


  
\bibitem[{\citenamefont{krug et~al.}(2003)\citenamefont{krug}}]{krug2003}
\bibinfo{author}{\bibfnamefont{H.-A}~\bibnamefont{Krug von Nidda}},
\bibinfo{author}{\bibfnamefont{R.}~\bibnamefont{Bulla}},
\bibinfo{author}{\bibfnamefont{N.}~\bibnamefont{Buttgen}},
\bibinfo{author}{\bibfnamefont{M.}~\bibnamefont{Heinrich}} \bibnamefont{and}
\bibinfo{author}{\bibfnamefont{A.}~\bibnamefont{Loidl}}, 
\bibinfo{journal}{Eur. Phys. J. B} \textbf{\bibinfo{volume}{34}},
\bibinfo{pages}{399} (\bibinfo{year}{2003}).



\bibitem[{\citenamefont{ikeda et~al.}(2000)\citenamefont{ikeda}}]{ikeda2000}
\bibinfo{author}{\bibfnamefont{Shin-Ichi}~\bibnamefont{Ikeda}},
\bibinfo{author}{\bibfnamefont{Yoshiteru}~\bibnamefont{Maeno}},
\bibinfo{author}{\bibfnamefont{Satoru}~\bibnamefont{Nakatsuji}},
\bibinfo{author}{\bibfnamefont{Masashi}~\bibnamefont{Kosaka}} \bibnamefont{and}
\bibinfo{author}{\bibfnamefont{Yoshiya}~\bibnamefont{Uwatoko}}, 
\bibinfo{journal}{Phys. Rev. B} \textbf{\bibinfo{volume}{62}},
\bibinfo{pages}{R6089} (\bibinfo{year}{2000}).




\bibitem[{\citenamefont{julian et~al.}(1999)\citenamefont{julian}}]{julian1999}
\bibinfo{author}{\bibfnamefont{S.R.}~\bibnamefont{Julian}},
\bibinfo{author}{\bibfnamefont{A.P.}~\bibnamefont{Mackenzie}},
\bibinfo{author}{\bibfnamefont{G.G.}~\bibnamefont{Lonzarich}},
\bibinfo{author}{\bibfnamefont{C.}~\bibnamefont{Bergemann}},
\bibinfo{author}{\bibfnamefont{R.K.W.}~\bibnamefont{Haselwimmer}},
\bibinfo{author}{\bibfnamefont{Y.}~\bibnamefont{Maeno}},
\bibinfo{author}{\bibfnamefont{S.}~\bibnamefont{NishiZaki}},
\bibinfo{author}{\bibfnamefont{A.W.}~\bibnamefont{Tyler}},
\bibinfo{author}{\bibfnamefont{S.}~\bibnamefont{Ikeda}} \bibnamefont{and}
\bibinfo{author}{\bibfnamefont{T.}~\bibnamefont{Fujita}}, 
\bibinfo{journal}{Physica B} \textbf{\bibinfo{volume}{259-261}},
\bibinfo{pages}{928} (\bibinfo{year}{1999}).





\end{thebibliography}
\end{document}